\documentclass[preprint,aps,showpacs,pre]{revtex4}
\usepackage[dvips]{graphicx}
\usepackage{color}

\begin{document}

\title{The Kramers-Kronig relations for usual and anomalous Poisson-Nernst-Planck models}
\author{Luiz Roberto Evangelista$^1$\footnote{Corresponding author. E-mail: lre@dfi.uem.br}, Ervin Kaminski Lenzi$^1$, and Giovanni Barbero$^{2,3}$}
\address{$^1$ Departamento de F\'{\i}sica, Universidade Estadual de Maring\'a\break
Avenida Colombo, 5790, 87020-900 Maring\'a, Paran\'a, Brazil\\
$^2$ Department of Applied Science and Technology,
Politecnico di Torino\\
 Corso Duca degli Abruzzi, 24, 10129 Torino, Italy\\
 $^3$ Turin Polytechnical University in Tashkent,
 17, Niyazov Str. Sobir Rakhimov district Tashkent, 100095 Uzbekistan}
\date{\today}

\begin{abstract}
The consistency of the frequency response predicted by a class of electrochemical impedance expressions is analytically checked by invoking the Kramers-Kronig (KK) relations. These expressions are obtained in the context of Poisson-Nernst-Planck usual (PNP) or anomalous (PNPA) diffusional models that satisfy Poisson's equation in a finite-length
situation.  The theoretical results, besides being successful in interpreting experimental data, are also shown to obey the KK relations when these relations are modified accordingly.
\end{abstract}

\pacs {68.43.Mn,66.10.C-,47.57.J-,47.57.E-,05.40.Fb}

\maketitle

The Kramers-Kronig (KK) relations are the appropriate tool to check the correctness of the frequency response of linear systems by assuring the construction of causal time-domain models~\cite{Toll,Kronig,Kramers,Bode}. These relations give a condition that is necessary and sufficient to verify if a given frequency response will yield a causal or a non-causal impulse response, without leaving the frequency domain. The KK relations have the origin in the Cauchy's theorem that provides the mathematical basis for causality~\cite{DDMac}. By using the notion of complex refractive index defined by means of an analytical continuation in the complex frequency plane, Kramers~\cite{Kramers} has shown that a signal cannot travel faster than $c$, the velocity of light in vacuo, in any medium for which the dispersion relation is satisfied~\cite{Toll}. More recently, KK relations have been intensively used to readily determine the validity of the data because they involve integral transform techniques that are independent of the physical process considered~\cite{DDMac}. This is particularly helpful in the field of the electrochemical impedance or immittance spectroscopy (EIS) applied to analyze experimental data of various systems. Indeed, EIS is a widely used technique for the electrical characterization of electrolytic cells by measuring the response of the material to an applied AC signal~\cite{Mac1}. At the impedance level, the set of experimental data is commonly analyzed by representing its frequency behavior in the complex plane. In this scenario, to obtain a physically acceptable interpretation of the measured response, besides using appropriate theoretical models, the KK relations have to be used to verify the consistency of the experimental data~\cite{Leur}. Thus, there is a general mathematical procedure which allows for the verification of the impedance data, i.e., they constitute conditions for ``good'' impedances that, in this manner, satisfy the criteria of linearity, causality, stability, and finiteness~\cite{Lasia}.

To interpret EIS experimental data from a theoretical perspective, in addition to the models based on equivalent circuits,  the small-signal Poisson-Nernst-Planck (PNP) diffusion model  is particularly important. Indeed, when one utilizes it to analyze immittance data, preferably using full complex nonlinear least squares (CNLS) fitting, as in the LEVMW computer program~\cite{Mac3,Mac4,Mac5}, it can lead to estimates of many more physically relevant electrical parameters than can any other available EIS model. The construction of any relevant theoretical model and, in particular, PNP or PNP modified models, has to be accompanied by checking their validity in  the terms
required by the KK relations.

In this Letter,  a set of recently proposed PNP models that are successful in interpreting the experimental EIS data \cite{lenzi,ciuchi,carlo,anca} is considered. It is demonstrated that they also obey the KK relations if these relations are slightly modified. These PNP models satisfy Poisson's equation in a finite length sample, e.g., a cell of thickness $d$ with flat electrodes of area $S$ placed at
the positions $z=\pm d/2$ of a Cartesian reference system. For simplicity's sake,  the analysis is limited to the case of full dissociation, of mobile charges with equal mobilities, in the presence of an electric field, of electrical potential $V$, due to the action of an
external power supply or to a charge separation. In this framework, the bulk densities, $n_p$ and $n_m$, where $p$ and $m$ stand for positive $+$ and negative $-$  ions, respectively,
are obtained by solving the standard form of the continuity equations, written as

\begin{equation}
\label{1}\frac{\partial n_p}{\partial t}=-\frac{\partial j_p}{\partial z} \quad {\rm and} \quad  \frac{\partial n_m}{\partial t}=-\frac{\partial j_m}{\partial z},
\end{equation}
in which the densities of currents are defined as

\begin{equation}
\label{1a}
j_{p,m} = - D_{p,m} \left(\frac{\partial n_{p,m}}{\partial z} \pm \frac{q N}{k_{\rm B}T} \frac{\partial V}{\partial z}\right),
\end{equation}
where $q$ is the electrical charge of the ions, $k_{\rm B}$ is the Boltzmann constant,  and $T$ the absolute temperature.
In addition, in thermodynamical equilibrium, the material contains a density of  $N$ positive and $N$ negative ions, per unit volume,
uniformly distributed across the sample.  The spatial profile of the electrical potential is governed by the Poisson's equation, in the form:

\begin{equation}
\label{1b}
\frac{\partial^2 V(z,t)}{\partial z^2} = -\frac{q}{\varepsilon}\left[n_p(z,t) - n_m(z,t)\right],
\end{equation}
where $\varepsilon$ is the dielectric coefficient of the medium.

This problem has been faced in the past~\cite{Mac11,Mac92} and
more recently~\cite{Ionescu,Libro,Mac252} for some significant boundary conditions and the details of the calculations for all these cases will be omitted here to save space. In the case of blocking electrodes, for which
\begin{equation}
j_{p,m} (\pm d/2, t) = 0,
\end{equation}
the electrical impedance is given by

\begin{equation}
\label{mamma1} Z=-i\,\frac{2}{\omega \varepsilon \beta^2 S}\left\{\frac{1}{\lambda^2 \beta} \tanh(\beta d/2)+i\frac{\omega d}{2 D}\right\},
\end{equation}
where $\omega$ is the frequency of the applied voltage,

\begin{equation}
\label{4-1}\beta=\frac{1}{\lambda}\sqrt{1+i\frac{\omega}{D}\lambda^2},
\end{equation}
and $\lambda= \sqrt{\varepsilon k_{\rm B}T/(2Nq^2)}$ is the Debye's screening length, and
$D_m=D_p=D$ is the diffusion coefficient assumed as the same for positive and negative ions.

The problem was also recently extended by assuming that the continuity equation contains a superposition of normal and fractional diffusion, the latter being characterized by a fractional coefficient $0 <\gamma \leq 1$, in the form:

\begin{equation}
\label{5}A\frac{\partial n_p}{\partial t}+B\frac{\partial ^\gamma n_p}{\partial t^{\gamma}}=D\frac{\partial}{\partial z}\left\{\frac{\partial n_p}{\partial z}+\frac{q N}{k_{\rm B} T}\frac{\partial V}{\partial z}\right\},
\end{equation}
where  $A$ is dimensionless, while the dimension of $B$ is $t^{\gamma-1}$.
The expression for the electrical impedance obtained in this case is~\cite{JPCM}

\begin{equation}
\label{mamma2} Z_e=-i\,\frac{2}{\omega \beta_e^2 S}\left\{\frac{1}{\lambda^2 \beta_e} \tanh(\beta_e d/2)+i\frac{\omega d}{2 D_e}\right\},
\end{equation}
where

\begin{equation}
\label{7}D_e=\frac{D}{A+B(i\omega)^{\gamma-1}}\quad {\rm and}\quad
\beta_e=\frac{1}{\lambda}\sqrt{1+i\frac{\omega}{D_e}\lambda^2}.
\end{equation}

A further generalization of the problem was achieved by considering  fractional time diffusion equations of distributed orders~\cite{Mainardi}. These equations may be formally written, for example, as

\begin{equation}
\label{eq2}
\int_{0}^{1}d\overline{\gamma} p(\overline{\gamma})\frac{\partial^{\overline{\gamma}}}{\partial t^{\overline{\gamma}}}n_{p,m}=-\frac{\partial}{\partial z}j_{p,m} (z,t),
\end{equation}
where $p(\gamma)$ is a distribution function of $\gamma$ and the fractional operator considered is the
Caputo one~\cite{Livro}, which can be defined as

\begin{eqnarray}
\frac{\partial^{\gamma}}{\partial t^{\gamma}}n_{p,m}(z,t)&\equiv&
_{\;\;\;\;t_{0}}^{\;\;\;\;\;{\cal{C}}}{D}_{t}^{\gamma}\left\{n_{p,m}(z,t)\right\}\nonumber \\&=&\frac{1}{\Gamma\left(k-\gamma\right)}
\int_{t_{0}}^{t}d\overline{t}\frac{n_{p,m}^{(k)}(z,\overline{t})}{(t-\overline{t})^{1-\gamma+k}}, \nonumber \\
\end{eqnarray}
with $0<k\leq 1$ and $n_{p,m}^{(k)}(z,\overline{t})$ representing the $k-{\rm th}$ derivative with respect to $\overline{t}$. As a particular case, it is useful to take the limit $t_0\rightarrow -\infty$
when one aims at studying the response of the system to a periodic applied potential as is done here~\cite{Livro}.
Note that Eq.~(\ref{eq2}) has the presence of fractional time operator of distributed order
which, depending on the choice of  $p(\gamma)$, can account for different diffusive regimes of the ions in the system, as will be discussed later. The order of these derivatives are consequently distributed according to the function $p(\gamma)$, that works as the weight factor for each regime (order). Thus, the general expression for the impedance is given by~\cite{JPCB}

\begin{equation}
\label{131}{Z}=-i\,\frac{2}{\omega \varepsilon \beta^2
{S}}\,\left\{\frac{1}{\lambda^2 \beta}\,\tanh\left(\frac{\beta
d}{2}\right)+\frac{ d}{2 {D}}{F}(i\omega)\right\}\,,
\end{equation}
where, now,

\begin{equation}
\label{beta} \beta=\frac{1}{\lambda}\,
\sqrt{1+{F}(i\omega)\,\frac{\lambda^2}{D}},
\end{equation}
with
\begin{eqnarray}
\label{efao}
{F}(i\omega)=\int_{0}^{1}d\overline{\gamma}p(\overline{\gamma})(i\omega)^{\overline{\gamma}}\;.
\end{eqnarray}
The presence of $F(i\omega)$ in Eqs.~(\ref{131}) and~(\ref{beta}) is responsible for the incorporation of an arbitrary number of diffusive regimes to the description of the diffusion of ions through the sample. In addition, it is noteworthy that the general expression for the impedance, Eq.~(\ref{131}), has exactly the same functional form of Eqs.~(\ref{mamma1}) and~(\ref{mamma2}),  which, in turn, can be hereafter faced as its particular cases.

Finally, to go one step further in the generalization process, one can consider again the fractional diffusion of distributed order governing the bulk behavior, but now  subjected to the boundary conditions~\cite{JCP}

\begin{eqnarray}
\label{e2}
\left. j_{\alpha}(z,t)\right|_{z=\pm\,\frac{d}{2}}\!=\!\left.\pm\int_{-\infty}^{t}d\overline{t}
\overline{\kappa}(t-\overline{t})\frac{d}{d\overline{t}}n_{\alpha}\!\left(z,\overline{t}\right)\right|_{z=\pm\,\frac{d}{2}},
\end{eqnarray}
where $\alpha = p, m$ and the right-hand term can be related to an adsorption-desorption process. In fact, for the specific choice of $\overline{\kappa}(t)=\kappa e^{-t/\tau}$,  we recover the adsorption-desorption processes at the surfaces governed by a kinetic equation that corresponds to the  Langmuir approximation~\cite{Libro}. Others choices of  $\overline{\kappa}(t)$  can be performed to incorporate memory effects and, consequently, non-Debye relaxation processes~\cite{Evangelista1}.
The impedance of the cell is

\begin{eqnarray}
\label{Impedance1}
Z\!\!&=&\frac{2}{i\omega \varepsilon {S}\alpha_{-}^{2}}\frac{\tanh\left(\alpha_{-}d/2\right)/(\lambda^{2}\alpha_{-})+d{C}/(2{D})}{1+
\overline{\kappa}(i\omega)\left(1+i\omega\lambda^{2}/D\right)\tanh \left(\alpha_{-}d/2\right)/(\lambda^{2}\alpha_{-})}
\end{eqnarray}
where $\alpha_{-}^{2}={F}\left(i \omega\right)/{D}+1/\lambda^{2}$ and $\alpha_{+}^{2}={F}\left(i \omega\right)/{D}$,
$\overline{\kappa}(i\omega)=e^{-i\omega t}\int_{-\infty}^{t}d\overline{t}\,\overline{\kappa}(t-\overline{t})e^{i\omega \overline{t}}$, and
$C={F}\left(i\omega\right)+i \alpha_{-}\omega\overline{\kappa}(i\omega)\tanh \left(\alpha_{-}d/2\right)$,

The presence of the kernel $\overline{\kappa}(t)$ in~(\ref{Impedance1}) gives to the electrical impedance a very general profile. This feature can be illustrated for two representative cases, among others. When one considers that $\overline{\tau}\left(\overline{\gamma}\right)=\delta(\overline{\gamma}-1)$, with $\overline{\kappa}(t)=\kappa e^{-t/\tau}$,   the case worked out in~\cite{Libro}, in which adsorption--desorption phenomena are incorporated to the analysis by means of a kinetic balance equation at the surfaces, is recovered.
Moreover, when $\tau(\overline{\gamma})=\delta(\overline{\gamma}-1)$,  with $\overline{\kappa}(t)=0$,  the usual form of the electrical impedance obtained in the situation of blocking electrodes is reobtained. Thus, the possible choices of the kernel allow one to handle
different expressions for the electrical impedance, suitable to face a large variety of experimental  situations.

In all the cases mentioned before, the system is governed by linear differential equations, of usual or fractional derivatives. Thus,  we expect that for the real ($R$) and imaginary ($X$) parts of $Z$ hold the KK relations. This is actually the case, as we can show by means of the simple calculation below. Notice, however, that these relations differ from those reported in Ref.~\cite{Mac1}, Eqs.~(21) and (23),  because the impedance diverges for $\omega \to 0$. To obtain the modified form of the KK relations,  we have just to consider the analytic function, in the upper-half part of the complex plane

\begin{equation}
\label{2} \oint_{\cal{C}}\frac{Z(\omega')}{\omega'-\omega}\,d\omega'={\cal P}\int_{-\infty}^{\infty} \frac{Z(\omega')}{\omega'-\omega}\,d\omega' -i\pi Z(\omega)-i \pi\lim_{\omega' \to 0}\omega'\frac{Z(\omega')}{\omega'-\omega},
\end{equation}
where $\cal{C}$ is a path in the complex plane, and $\cal{P}$ denotes the principal value of the integral, on the real axis. The third contribution on the LHS, i.e.

\begin{equation}
\label{3}-i \pi\lim_{\omega' \to 0}\omega'\frac{Z(\omega')}{\omega'-\omega},
\end{equation}
is necessary because $Z(\omega)$ diverges for $\omega=0$. Since $Z$ is analytic in ${\cal C}$, from Eq.~(\ref{2}) we get

\begin{equation}
\label{4}Z(\omega)=i\frac{\pi}{\omega}\lim_{\omega'\to 0}\omega' Z(\omega')+{\cal P}\int_{-\infty}^{\infty} \frac{Z(\omega')}{\omega'-\omega}\,d\omega'.
\end{equation}
Considering that $Z=R+i X$, from  Eq.~(\ref{4}),  we obtain the KK relations as

\begin{eqnarray}
\label{5}R(\omega)&=&\frac{1}{\omega} {\cal L}_R+\frac{1}{\pi}{\cal P}\int_{-\infty}^{\infty} \frac{X(\omega')}{\omega'-\omega}\,d\omega' \quad {\rm and}\\
\label{6}
X(\omega)&=&\frac{1}{\omega} {\cal L}_X-\frac{1}{\pi}{\cal P}\int_{-\infty}^{\infty} \frac{R(\omega')}{\omega'-\omega}\,d\omega',
\end{eqnarray}
where

\begin{equation}
\label{7}{\cal L}_R=\lim_{\omega'\to 0}\omega' R(\omega') \quad{\rm and}\quad {\cal L}_X=\lim_{\omega'\to 0}\omega' X(\omega').
\end{equation}
For $\gamma <1$, we have

\begin{equation}
\label{8}{\cal L}_R=\lim_{\omega'\to 0}\omega' R(\omega')=0 \quad{\rm and}\quad{\cal L}_X=\lim_{\omega'\to 0}\omega' X(\omega')={\cal L},
\end{equation}
where $\cal{L}$ is a finite quantity. Thus, the KK relations for the class of problems we are considering are

\begin{eqnarray}
\label{9}R(\omega)&=&\frac{1}{\pi}{\cal P}\int_{-\infty}^{\infty} \frac{X(\omega')}{\omega'-\omega}\,d\omega' \quad {\rm and} \\
\label{10}
X(\omega)&=&\frac{1}{\omega} {\cal L}-\frac{1}{\pi}{\cal P}\int_{-\infty}^{\infty} \frac{R(\omega')}{\omega'-\omega}\,d\omega',
\end{eqnarray}
which can be conveniently rewritten, respectively, as 

\begin{eqnarray}
\label{11}R(\omega)&=&\frac{2}{\pi}{\cal P}\int_{0}^{\infty} \frac{\omega' X(\omega')-\omega X(\omega)}{\omega'^2-\omega^2}\,d\omega' \quad {\rm and}\\
\label{12}
X(\omega)&=&\frac{1}{\omega} {\cal L}-2\frac{\omega}{\pi}{\cal P}\int_{0}^{\infty} \frac{R(\omega')-R(\omega)}{\omega'^2-\omega^2}\,d\omega'.
\end{eqnarray}
We are now ready to apply this formalism to the electrical impedance expressions presented before.

 Consider first Eq.~(\ref{131}), whose low frequency limit is given by the expression
 
\begin{eqnarray}
\label{Approx1}
Z\approx \frac{2\lambda}{i \omega\varepsilon S}\left[1+\frac{d}{2D}F(i\omega)\right]
\end{eqnarray}
and, consequently,

\begin{equation}
\label{8-1}{\cal L}=\frac{2 \lambda}{\varepsilon S},
\end{equation}
because, from Eq.~(\ref{efao}), it follows that
$\lim_{\omega \to 0} F(i\omega)=0$, for $0< \gamma \leq 1$. 
\begin{figure}[!t]
\centering
\includegraphics[scale=0.3]{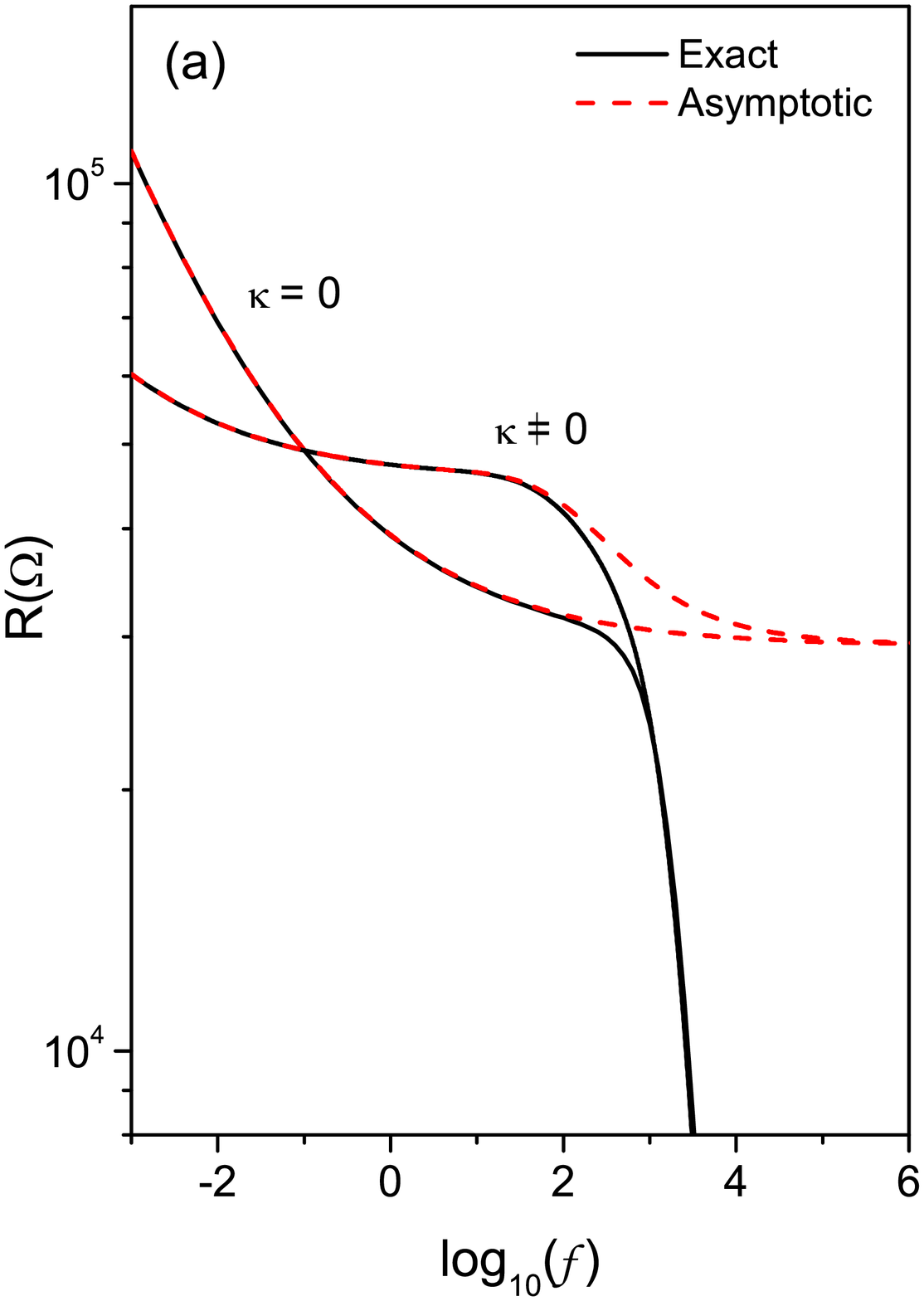}
\includegraphics[scale=0.3]{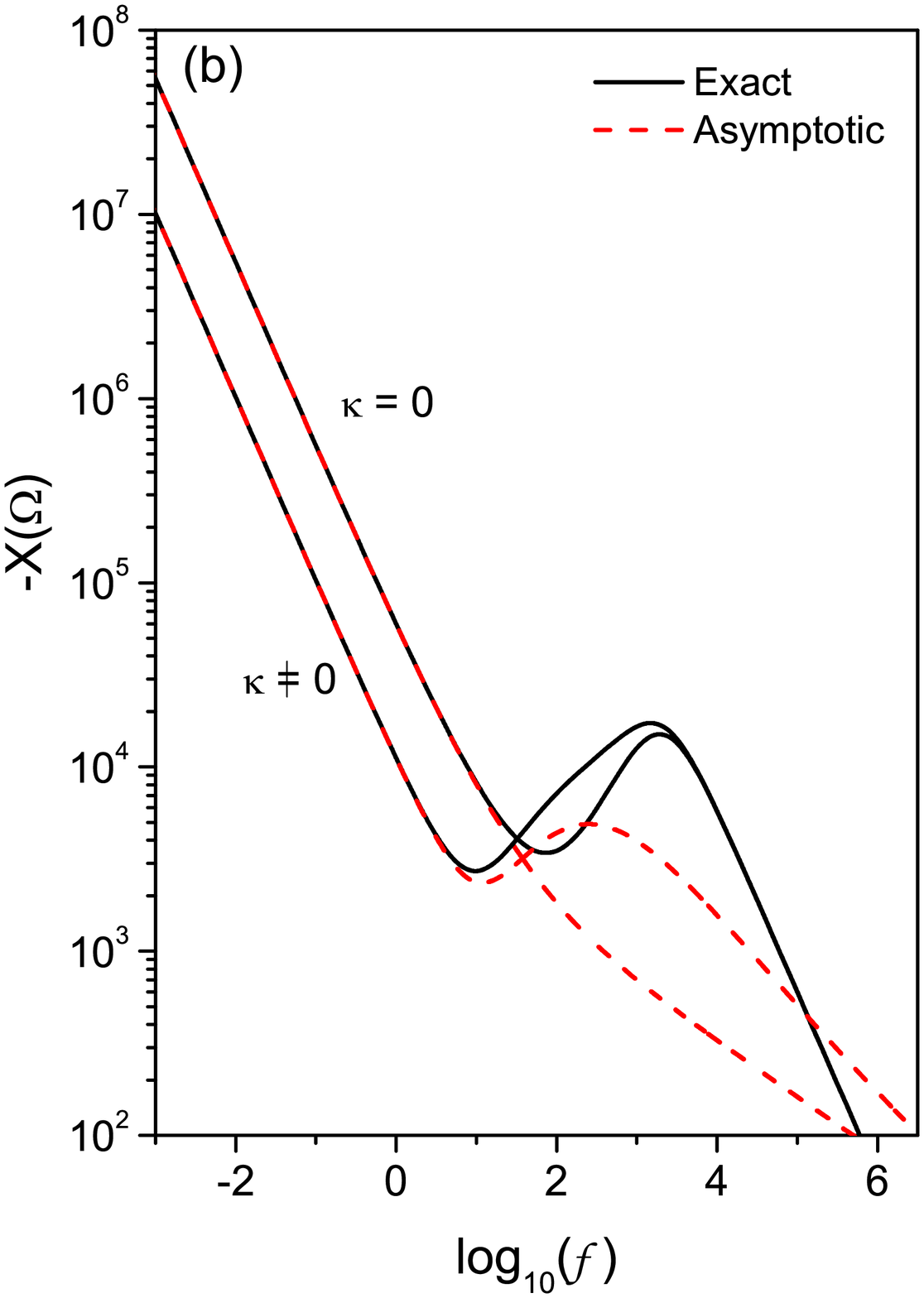}
\caption{Behavior of the real, ${\mbox{R}}={\mbox{Re}}\,Z$ (a), and imaginary, ${\mbox{X}}={\mbox{Im}}\,Z$ (b), parts of the impedance versus the frequency for $\kappa\neq0 $ and $\kappa=0$. The black solid line corresponds to the exact case and the dashed red line is the asymptotic behavior. The curves have been drawn for the following
values of the parameters relative to a liquid-crystalline system:  $S=2\times10^{-3}\;{\rm m}^{2}$, $\varepsilon= 7.5\,\varepsilon_{0}$ ($\varepsilon_{0}= 8.85\times 10^{-12}\; {\rm C}^{2}/({\rm Nm}^{2})$), $\gamma=0.7$, $D=4\times 10^{-12}\;{\rm m}^{2}/{\rm s}$, $d=50\times10^{-6}{\rm m}$,  $q=1.6\times 10^{-19} {\rm C}$, $\kappa=10^{-5}\;{\rm m/s}$, $\tau=0.01\, {\rm s}$~\cite{Maximus,DHave}, $A=0.6$, $B=0.4$, and $\lambda= 2.27 \times 10^{-8} {\rm m}$.}
\label{Figure1}
\end{figure}
This quantity coincides with the inverse of the total capacitance of the sample, in the dc limit. Indeed, it coincides with the inverse of the resulting series capacitance of two 
equal capacitances   $C_D=\varepsilon S/\lambda$, that can be identify with 
the capacitance of the surface layer, i.e., the Gouy-Chapman double layer capacitance. 
It is then clear that the entire class of expressions of the kind considered in Eq.~(\ref{131}), for which $\lim_{\omega \to 0} F(i\omega) \to 0$, obeys the modified KK relations and, as expected,  are good candidates as theoretical tools for interpreting EIS experimental data.

Consider now Eq.~(\ref{Impedance1}), for the 
 case characterized by $\overline{\kappa}(t)=\kappa e^{-t/\tau}$, i.e., $\overline{\kappa}(i\omega)=\kappa\tau/(1+i\omega\tau)$, which, as mentioned before, represents, in the frequency domain, the usual kinetic equation connected with adsorption-desorption phenomenon (Langmuir's approximation) in the time domain. It is possible to show that the low frequency limit of Eq.~(\ref{Impedance1}) is given by
 
\begin{eqnarray}
\label{Approx2}
Z\approx \frac{2\lambda}{i \omega\varepsilon S}\frac{\lambda}{\lambda+\overline{\kappa}(i\omega)}\left\{1+\frac{d\lambda}{2D}\left[F(i\omega)+i\omega\overline{\kappa}(i\omega)
\right]\right\}
\end{eqnarray}
which, by means of Eq.~(\ref{8}), allows one to obtain

\begin{equation}
\label{8-2}{\cal L}=\frac{2 \lambda}{\varepsilon S}\frac{1}{1+\kappa\tau/\lambda}.
\end{equation}
In Fig.~(\ref{Figure1}), the exact results, Eqs.~(\ref{131}) and~(\ref{Impedance1}), and the approximated ones, i.e., Eqs.~(\ref{Approx1}) and~(\ref{Approx2}), are illustrated for the cases discussed above in order to compare their low frequency behavior. 

It is worth mentioning that similarly to the previous
result obtained, i.e., for Eq.~(\ref{8-1}), Eq.~(\ref{8-2}) is also 
connected with the inverse of the total capacitance of the
sample, but now taking into account the influence of the
adsorption process occurring at the surfaces of the electrodes. 
Thus, in the case in which the adsorption process is present, the effective thickness intervening in the capacitance of the double-layer is the sum of the Debye's screening
length with the quantity $ \kappa \tau$, which has dimensions of length. 
In a phenomenological perspective,
it represents an effective thickness of the layer over which
the adsorption--desorption phenomenon takes place near
the interface. Again, the KK relations are obeyed by the general expression, Eq.~(\ref{Impedance1}) for a large class of choices for 
$ {\overline{\kappa}} (i\omega)$ that remains finite in
the low frequency domain. These choices have their
counterparts in the time--domain which, in general, is
more helpful to interpret the phenomena occurring near
the interface in terms of boundary conditions. This feature is remarkable illustrated in the
particular case analyzed above (Langmuir's approxi-
mation). In this example, the emergence of 
an adsorbing layer was automatically incorporated to the
resulting double-layer capacitance, renormalizing it and
allowing for a simple interpretation of the phenomenological 
parameters entering the kinetic equation. Notwithstanding, even
if obtained for a particular case, this noticeable result
permits one to expect that the resulting effects of other
significant phenomena, occurring near to the electrodes,
could be incorporated to the description of the global
properties of the electrolytic cell in the PNP or PNPA
models.

To summarize,  the consistency of
the frequency response of a class of electrochemical
impedance expressions, obtained in the ambit of
usual or anomalous (PNP or PNPA) models, as well as some of their
generalizations that takes into account different types of
boundary conditions,  have been theoretically analysed. 
The slightly modified expressions
of the KK relations presented here, and obeyed by these
expressions, can be also used to verify the correctness of
the impedance date obtained by means of the EIS 
technique in the field of electro-chemistry and condensed 
matter physics.

\section*{Acknowledgments}

This work was partially supported by the Brazilian Agency, CNPq, by means of the National Institutes of Science and Technology of Complex Fluids -- INCT-FCx (L. R. E.) and Complex Systems -- INCT-SC (E. K. L.).

\end{document}